\documentclass[reqno]{amsart}
\usepackage{xcolor}
\usepackage[utf8]{inputenc}
\usepackage{hyperref}
\usepackage{enumitem}
\usepackage{caption}
\usepackage{a4wide,amsmath}
\usepackage{amssymb}
\usepackage{mathtools}
\usepackage{mathrsfs}
\usepackage{amsthm}
\numberwithin{equation}{section}
\usepackage{tikz}
\usepackage{pgfplots}
\theoremstyle{plain}

\theoremstyle{definition}

\theoremstyle{remark}

\renewcommand{\Im}{\mathrm{Im}}

\renewcommand{\d}{\;\mathrm{d}}

\begin{document}

\title{Optimal profile design for acoustic black holes using Timoshenko beam theory}

\author{Kasper S. S{\o}rensen}
\email{kasper@math.aau.dk}
\address{Department of Mathematical Sciences, Aalborg University \\ Skjernvej 4A, 9220 Aalborg {\O}, Denmark.}

\author{Horia D. Cornean}
\email{cornean@math.aau.dk}
\address{Department of Mathematical Sciences, Aalborg University \\ Skjernvej 4A, 9220 Aalborg {\O}, Denmark.}

\author{Sergey Sorokin}
\email{svs@mp.aau.dk}
 \address{Department of Materials and Production, Aalborg University \\ Fibigerstr{\ae}de 16, 9220 Aalborg {\O}, Denmark.}

\begin{abstract}
We revisit the problem of constructing one-dimensional acoustic black holes. Instead of considering the Euler-Bernoulli beam theory, we use Timoshenko's approach instead, which is known to be more realistic at higher frequencies. Our goal is to minimize the reflection coefficient under a  constraint imposed on the normalized wave number variation. We use the calculus of variations in order to derive the corresponding Euler-Lagrange equation analytically and then use numerical methods to solve this equation in order to find the optimal height profile for different frequencies. We then compare these profiles to the corresponding ones previously found using the Euler-Bernoulli beam theory and see that in the lower range of the dimensionless frequency $\Omega$ (defined using the largest height of the plate), the optimal profiles almost coincide, as expected. For higher such frequencies, even for values where Euler-Bernoulli theory should still be marginally valid, the profiles predicted using Euler-Bernoulli differ substantially from the correct ones predicted by Timoshenko theory. One explanation for this phenomenon is that unlike in the constant height case, in our setting the wave numbers also depend on the ratio between the smallest and the largest heights.
\end{abstract}

\maketitle

%\KEY
%ABH
%\EndKEY
%
%\MSC
%NaN
%\EndMSC

%
%
%
%
%
%
%
%
%INTRODUCTION
%
%
%
%
%
%
\section{Introduction}
The study of acoustic black holes originates from the seminal work of Mironov \cite{Mironov1988}. He showed, using analytic methods, that if one can construct a plate with an ideal wedge (i.e. the thickness of the wedge goes smoothly to zero in a finite interval (see Fig.~\ref{fig:wedgetozero})), then the velocity of the flexural wave moving towards the tip would go to zero and thus never reach the end of the rod. This is equivalent to saying that no reflection of the flexural wave would occur and hence an acoustic black hole has been created. 

Unfortunately, it is not possible to create an ideal wedge in practice, thus the current research concerning acoustic black holes is focused around minimizing the reflection of the wave when it hits the end of the wedge. Acoustic black holes for other geometries and higher dimensions have also been considered, see eg. \cite{LeeJeon2017} and \cite{GuaschEtAl2017} for different geometries and \cite{Krylov2007} for the  2-dimensional case. Several different directions have been studied in connection to this minimization problem, e.g. profile optimization by adding a thin dampening layer on the wedge. To the best of the authors knowledge, the study of acoustic black holes has, up until now, only been considered using the Euler-Bernoulli beam theory. We refer the reader to the recent survey paper \cite{PelatEtAl2020} and the references therein for a thorough overview of the state of-the-art in this field.

In our paper, inspired by some recent work \cite{StottrupEtAl2021}, we apply the mathematical theory of calculus of variations in order to determine an optimal wedge profile which minimizes the reflection of flexural waves. For the first time regarding acoustic black holes, we consider waves using the Timoshenko beam theory. Timoshenko's approach can be considered as an extension of the one of Euler-Bernoulli, in the sense that the their predictions almost coincide for dimensionless frequencies (defined in \eqref{hc1} where $h_1$ is the largest possible height of the plate) up to $\Omega \approx 0.3$, see e.g. \cite{Miklovitz1980} and \cite{SorokinChapman2011}. 
 
For dimensionless frequencies higher than this value, it is known that the Euler-Bernoulli beam theory starts deviating from the true one, while the Timoshenko  approach is known (from the comparison with the exact solution of the Rayleigh-Lamb problem \cite{Miklovitz1980},\cite{Achenbach1973}) to hold true up to $\Omega \approx 3.5$ (see eg. \cite{StephenPuchegger2006}). This is in part due to the fact the the Timoshenko beam theory takes into account shear deformation along with rotational bending effects, which are neglected in the Euler-Bernoulli approximation. 

We stress that the above values for $\Omega$ are derived for a {\it constant} height of the plate. Because we let the height of the plate to vary, we will see that the wave numbers also depend on the ratio between the smallest and the largest heights. This means that in our setting, $\Omega$ is no longer the only parameter which decides the transition between the two regimes. 

Due to this much higher validity range of the Timoshenko approach, we find it natural to investigate the possibility of constructing acoustic black holes within this more complete theory. Moreover, it is also quite relevant to compare our results with the earlier ones based on Euler-Bernoulli, and to find the frequency threshold above which the optimal profiles predicted by two approaches start to substantially differ from each other. 

The Timoshenko beam theory also predicts the appearance of a second wave. This second wave is considered analytically in our paper, but we have not yet been able to identify real materials where the second wave appears in an acoustic  frequency range.  Thus the numerical investigations in this paper are only with regard to the first wave. A numerical investigation of the second wave, along with trying to find some material (or meta-material) where the second wave appears in the acoustic frequency range, is very interesting but it will be postponed for a future work.    

\subsection{Problem statement}
The primary aim of this paper is to determine a height profile for a truncated plate (see Fig.~\ref{fig:trunc}), which minimizes the reflection of a flexural wave, when the wave propagation is considered within Timoshenko's beam theory. The system of partial differential equations describing the vibration of a Timoshenko beam is given by (see \cite{Miklovitz1980} for a derivation)
\begin{equation}\label{eq:timoshenkodiffeq}
\begin{aligned}
-\rho A \frac{\partial^2w}{\partial \tau^2}(x,\tau) + \kappa GA\Big(\frac{\partial^2 w}{\partial x^2}(x,\tau) - \frac{\partial \psi}{\partial x}(x,\tau)\Big) + q(x,\tau) &= 0, \\
-\rho I \frac{\partial^2\psi}{\partial \tau^2}(x,\tau) + EI \frac{\partial^2\psi}{\partial x^2}(x,\tau) + \kappa GA\Big(\frac{\partial w}{\partial x}(x,\tau) - \psi(x,\tau) \Big) &= 0,
\end{aligned}
\end{equation}
where $\rho$ is the density of the material, $A=Bh_d(x)$ is the cross section area where $B$ is the constant width, while $h_d(x)$ is the height, $E = E_0(1-i\eta)$ is the complex elastic modulus with loss $\eta>0$, $G=\tfrac{E}{2(1+\nu)}$ is the shear modulus with $\nu$ its Poisson ratio, $I=\tfrac{B(h_d(x))^3}{12}$ is the second moment area, $\kappa$ is the Timoshenko shear coefficient and $q$ is the distributed load.
\begin{figure}[!h]
	\begin{minipage}{0.49\textwidth}
		\begin{tikzpicture}
		\begin{axis}[xmin=0,xmax=2/3,ymin=-2/3,ymax=2/3, xtick={0,1/2},height=0.6\textwidth,width=\textwidth,xticklabels={0, $x_1$},ytick={0,1/2,-1/2},yticklabels={0,$h_1$,$-h_1$}]
		\addplot[thick,black,domain=0:1/2] {2*x^2};
		\addplot[thick,black,domain=1/2:2/3] {1/2};
		\addplot[thick,black,domain=0:1/2] {-2*x^2};
		\addplot[thick,black,domain=1/2:2/3] {-1/2};
		\end{axis}
		\end{tikzpicture}
		\captionof{figure}{Non-truncated wedge}  \label{fig:wedgetozero}
	\end{minipage}
	\begin{minipage}{0.49\textwidth}
		\begin{tikzpicture}
		\begin{axis}[xmin=0,xmax=2/3,ymin=-2/3,ymax=2/3, xtick={1/8,1/2},height=0.6\textwidth,width=\textwidth,xticklabels={$x_0$, $x_1$},ytick={1/32,1/2},yticklabels={$ h_0 $,$h_1$}]
		\addplot[thick,black,domain=1/8:1/2] {2*x^2};
		\addplot[thick,black,domain=1/2:2/3] {1/2};
		\addplot[thick,black,domain=1/8:1/2] {-2*x^2};
		\addplot[thick,black,domain=1/2:2/3] {-1/2};
		\addplot[thick,black] coordinates{(1/8,-1/32) (1/8,1/32)};
 		\node at (axis cs:0.3,0.38) {\footnotesize $h_d(x)$};
		\end{axis}
		\end{tikzpicture}
		\captionof{figure}{Truncated wedge}\label{fig:trunc}
	\end{minipage}
\end{figure}

We will see in the next section that when looking for propagating solutions with an angular frequency $\omega$ to the "locally homogeneous" system  \eqref{eq:timoshenkodiffeq} (i.e. with $q=0$), the "local" wave number $k_d$ ("$d$" from dimensional) must depend on the profile through the height $h$ of the profile. 

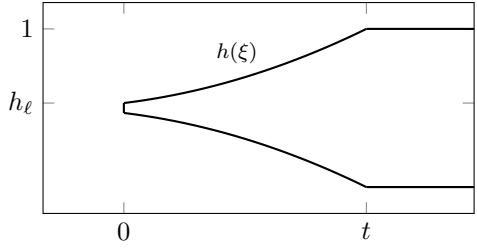
\begin{figure}[!h]
	\begin{minipage}{0.49\textwidth}
		\centering
		\begin{tikzpicture}
		\begin{axis}[xmin=0,xmax=2/3,ymin=-2/3,ymax=2/3, xtick={1/8,1/2},height=0.6\textwidth,width=\textwidth,xticklabels={0, $t$},ytick={1/32,1/2},yticklabels={$ h_\ell $,$1$}]
		\addplot[thick,black,domain=1/8:1/2] {2*x^2};
		\addplot[thick,black,domain=1/2:2/3] {1/2};
		\addplot[thick,black,domain=1/8:1/2] {-2*x^2};
		\addplot[thick,black,domain=1/2:2/3] {-1/2};
		\addplot[thick,black] coordinates{(1/8,-1/32) (1/8,1/32)};
		\node at (axis cs:0.3,0.35) {\footnotesize $h(\xi)$};
		\end{axis}
		\end{tikzpicture}
		
	\end{minipage}
	\caption{Dimensionless,  truncated wedge}
	\label{fig:nondimtrunc}
\end{figure}

We also denote with  $c=\sqrt{\tfrac{E}{\rho}}$ the P-wave propagation speed, and with  $\tilde{W}$ the  amplitude of the lateral displacement $w(x,t)$. We define some  dimensionless variables as follows (see Fig.~\ref{fig:nondimtrunc}):
\begin{align}\label{hc1}
h_\ell &= \frac{h_0}{h_1}, \quad  t = \frac{x_1-x_0}{h_1}, \quad 0\leq \xi\leq t,\quad \Omega = \frac{\omega h_1}{c}, \quad W = \frac{\tilde{W}}{h_1},\\
& h(\xi)=\frac{h_d(x_0+\xi h_1)}{h_1}, \quad k(\xi) = k_d(x_0+\xi h_1)h_1. \nonumber
\end{align}
 
In order to quantify the performance of an acoustic black hole, we use the reflection coefficient (see eg. \cite{Krylov2004} or \cite{Mironov1988})
\begin{align}\label{eq:reflectioncoeff}
R=\exp\Big(-2\int_{0}^{t} \Im(k(\xi)) \d \xi \Big),
\end{align}
a formula whose validity is based on a slow variation assumption \cite{KrylovTilman2004}:
\begin{align}\label{eq:NWV}
\Big\vert \frac{1}{k^2}\frac{\d k}{\d \xi} \Big\vert \ll 1.
\end{align}
We refer to the left-hand side of \eqref{eq:NWV} as the normalized wave number variation. In practice, a normalized wave number variation less than $0.3$ is often seen as reasonable \cite{FeurtadoConlon}.

To summarize, given $k(\xi)$ as a function of the profile (to be found in the next section),  our aim is to find a height function $h \colon [0,t] \to [h_\ell,1]$ (see Fig~\ref{fig:nondimtrunc}) which satisfies:
\begin{enumerate}[label=(\roman*)]
\item $h \in C^1$, i.e.\ differentiable with a continuous derivative,
\item $h(0)=h_\ell$, $h(t)=1$, and $h$ is non-decreasing,
\item $h$ is a minimizer of \eqref{eq:reflectioncoeff} under the constraint \eqref{eq:NWV}; this statement will be made mathematically more precise later. 
\end{enumerate}
Note, that $h$ being a minimizer of \eqref{eq:reflectioncoeff} is equivalent to $h$ being a maximizer of
\begin{align}\label{eq:integralrecflectioncoeff}
\int_0^t \Im (k(\xi)) \d \xi.
\end{align}

\subsection{Structure of the paper}
The rest of this paper is as follows. In Section \ref{sec:approach} we develop an analytic framework in order to maximize \eqref{eq:integralrecflectioncoeff} under some further conditions. This framework is based on the calculus of variations and we formulate an Euler-Lagrange equation which must be satisfied by a maximizer. 

Section \ref{sec:numerics} is dedicated to the numerical implementation of the analytic results from the first part. At the end, we summarize our results and formulate a number of open questions which we believe to be worth considering in the near future.

\section{The analytic results}\label{sec:approach}
Our first task is to derive an expression for the "local" wave numbers, reasoning like in the WKB approximation under a slow variation assumption. Namely, we look for a solution of  \eqref{eq:timoshenkodiffeq} when $q=0$ of the form
\begin{align*}
w(x,t)=W(x)\; h_1\; e^{ik_d(x)x - i \omega t} \qquad \textup{ and } \qquad \psi(x,t)=\Psi(x) e^{ik_d(x)x-i\omega t}.
\end{align*}
The zeroth-order component in the WKB-like expansion leads to the following system of linear equations in $W$ and $\Psi$:
\begin{align*}
&\Big(-k^2+\frac{2(1+\nu)}{\kappa}\Omega^2\Big)W - ik\Psi = 0, \\
&\frac{6ik\kappa }{(1+\nu)(h(\xi))^2}W+\Big(-k^2 + \Omega^2 - \frac{6\kappa }{(1+\nu)(h(\xi))^2}\Big)\Psi =0.
\end{align*}
In order to get a non-trivial solution for the pair $W$ and $\Psi$, the coefficient matrix of the above linear system must be singular, hence its determinant must vanish. This gives the Timoshenko dispersion relation
\begin{align}\label{eq:timoshenkodispersioneq}
k^4 - \Big(1+ \frac{2(1+\nu)}{\kappa}\Big)k^2\Omega^2 - \frac{12 }{(h(\xi))^2} \Omega^2 + \frac{2(1+\nu)}{\kappa}\Omega^4 = 0.
\end{align}
Next we want to find the solutions to this equation. To do so, we observe that the left-hand side is an ordinary second order polynomial with $k^2$ as variable. In our paper the complex square root is defined as follows: $z=re^{i\phi}$ with $r>0$ and $0\leq \phi<2\pi$, then   $\sqrt{z}=\sqrt{r}\; e^{i\phi/2}$. 

 Even though there are four algebraic solutions, we are only interested in those which remain bounded for $x\geq 0$. This demands that the imaginary part of the physical solutions is non-negative. Thus the solution(s) are among the expressions below:
\begin{align}\label{eq:timoshenkowavenumber}
k(\xi)=\pm \frac{\sqrt{(\kappa+2(1+\nu))\Omega^2 \pm \sqrt{(\kappa -2(1+\nu))^2\Omega^4 + \frac{48 \kappa^2}{h(\xi)^2} \Omega^2}}}{\sqrt{2\kappa}},
\end{align}
with the extra condition  ${\rm Im}(k)\geq 0$. On the other hand, we are neither interested in evanescent waves where the imaginary part is large. That is why it is important to analyze how the imaginary part of these solutions behaves as a function of $\Omega$. We can already see from the above formula that, by our choice of scaling, the term $\frac{48 \kappa^2}{h(\xi)^2} \Omega^2$ can become very large when $h(\xi)$ gets very small, hence the transition between Euler-Bernoulli and Timoshenko is no longer uniquely determined by $\Omega$, but also by the ratio $h_\ell$ between the smallest and largest values of the height. 

\subsection{Frequency ranges}\label{sec:freqrange}
We now make a short detour in order to study the behaviour of the wave numbers in different frequency ranges. The dimensionless frequency $\Omega$ has -by construction- an infinitesimal positive imaginary part, which can be traced back to the negative imaginary part of the elastic modulus $E$; the same holds true for $\Omega^2$ and $\Omega^4$. It is very important to warn the reader that in our setting, the transition between Euler-Bernoulli and Timoshenko theories is not solely determined by $\Omega$, but also by $h_\ell$.

We start our analysis in the regime  $|\Omega|\ll 1$, where Euler-Bernoulli's and  Timoshenko's beam theories should agree (see \cite{Miklovitz1980} and \cite{SorokinChapman2011}). Let us confirm this. In this frequency regime, we have $|\Omega|^4\ll |\Omega|^2$, hence the inner square root in  
\eqref{eq:timoshenkowavenumber} "almost" equals  $\frac{\sqrt{48}\kappa}{h(\xi)}\Omega$. This linear term is also much larger in absolute value than $|\Omega|^2$, hence:
\begin{align*}
k(\xi) \approx \pm \frac{\sqrt[4]{12}}{\sqrt{h(\xi)}}\sqrt{\pm \Omega}.
\end{align*}
 The solution with $\sqrt{-\Omega}$ is evanescent, hence the only solution with a non-negative small imaginary part is the one with two $+$ signs, which coincides with the Euler-Bernoulli solution.  

In other words, we just showed that the solution
\begin{align}\label{EB}
k_+(\xi) =\frac{\sqrt{(\kappa+2(1+\nu))\Omega^2 + \sqrt{(\kappa -2(1+\nu))^2\Omega^4 + \frac{48 \kappa^2}{(h(\xi))^2} \Omega^2}}}{\sqrt{2\kappa}}
\end{align}
is the one which recovers the Euler-Bernoulli wavenumber for small frequencies. When the frequency grows, the term with $\Omega^4$ becomes dominant and there exists a critical frequency $\Omega_{\mathrm{crit}}$, such that when $|\Omega |> \Omega_{\mathrm{crit}}$, the wave behaves like a Rayleigh wave, i.e.
$$
k_+ \approx  \sqrt{\frac{|\kappa-2(1+\nu)|}{2\kappa}}\; \Omega.
$$

This means that for very high frequencies, the shape profile will have no impact on this wave. Note, that $\Omega_{\mathrm{crit}}$ can be chosen to be the solution to the equation $(\kappa-2(1+\nu))^2\Omega_{\mathrm{crit}}^2 = \frac{48\kappa^2}{h_\ell ^2}$.

On the other hand, when $|\Omega|$ grows, there exists a second wave number which obeys our two conditions, namely: 
\begin{align*}
k_-(\xi) = \frac{\sqrt{(\kappa+2(1+\nu))\Omega^2 - \sqrt{(\kappa -2(1+\nu))^2\Omega^4 + \frac{48 \kappa^2}{(h(\xi))^2} \Omega^2}}}{\sqrt{2\kappa}}.
\end{align*}
which appears when $|\Omega| > \Omega_{\mathrm{cut-on}}$, where $\Omega_{\mathrm{cut-on}}$ solves $8\kappa(1+\nu)\Omega^2 = 48\kappa^2h_\ell ^{-2}$. In this frequency range the behaviour of the second wave number is given as
\begin{align}\label{hc7}
k_-(\xi) \approx |\kappa-2(1+\nu)|^{-1/2}\Big(8\kappa(1+\nu)\Omega^2-\frac{48\kappa^2}{(h(\xi))^2} \Big)^{1/2}.
\end{align}
It should be noted that in general, it is expected that the validity of the Timoshenko beam theory for the second wave only holds ``close'' to $\Omega_{\mathrm{cut-on}}$.
%{\color{blue}(Note that $\Omega_{\mathrm{min}}=1.9764$, which coincides with the value in Sergeys notes but for a frequency far higher than the acustic frequency range)}.

\subsection{The imaginary part of the wave number}\label{sec:imaginarypart}
We now return to solving the minimization problem. To get an expression for the imaginary part of the wave numbers, we make a first order approximation of \eqref{eq:timoshenkowavenumber} with respect to $\eta$. This leads to
\begin{align}\label{hc3}
k(\xi) \approx \pm\Big(\frac{\omega^2 h_1^2}{2c_0^2\kappa}\Big)^{1/2} \sqrt{\kappa^{+}(1+i\eta) \pm \sqrt{(\kappa^{-})^2 + \tilde{h}_\xi + i \eta \Big(2(\kappa^{-})^2 + \tilde{h}_\xi  \Big)}},
\end{align}
where $c_0^2=\tfrac{E_0}{\rho}$, $\kappa^\pm = \kappa \pm 2(1+\nu)$ and $\tilde{h}_\xi=\tfrac{48\kappa^2c_0^2}{(h(\xi))^2\omega^2h_1^2}$. It has the form $\sqrt{a+bi \pm \sqrt{c+di}}$, where $\vert b \vert, \vert d \vert \ll 1$ and $c > 0$. By a first order Taylor approximation we can write this as 
\begin{align*}
\sqrt{a+bi \pm \sqrt{c+di}} \approx \sqrt{a \pm \sqrt{c}}+i \frac{b \pm \frac{d}{2\sqrt{c}}}{2\sqrt{a \pm \sqrt{c}}},
\end{align*}
when $\kappa^+ \pm \sqrt{(\kappa^-)^2+\tilde{h}_\xi}>0$. Thus under this condition, we can approximate the non-negative imaginary part of the two solutions by
\begin{align*}
\Im (k_\pm) \approx \eta \Big(\frac{\omega^2h_1^2}{8c_0^2\kappa} \Big)^{1/2} \frac{\kappa^+ \pm \frac{2(\kappa^-)^2 + \tilde{h}_\xi}{2\sqrt{(\kappa^-)^2 + \tilde{h}_\xi}}}{\sqrt{\kappa^{+} \pm \sqrt{(\kappa^-)^2 + \tilde{h}_\xi}}}.
\end{align*}
\subsection{Dampening layer}\label{sec:damplayer}
Before we continue with the method of maximizing this integral, we make a comment about attaching an absorbing layer to the wedge surfaces. Here we follow the ideas from \cite{Krylov2004} and \cite{KrylovTilman2004}.

By attaching a visco-elastic layer of constant thickness $\delta$ to one of the sides of a plate of constant thickness $h$, we get the following additional loss factor
\begin{align*}
\Xi = \frac{\tilde{\eta}}{1+\big(\alpha_2\beta_2(\alpha_2^2+12\alpha_{21}^2)\big)^{-1}},
\end{align*} 
where $\tilde{\eta}$ is the loss factor of the visco-elastic layer, $\alpha_2=\delta/h$, $\beta_2=E_1/E_0$, where $E_0$ is Young's modulus for the plate and $E_1$ is Young's modulus for the visco-elastic layer and $\alpha_{21}=(1+\alpha_2)/2$. This formula has been derived under the assumption $(\delta/h)(E_1/E_0) \ll 1$. 

If we instead cover both sides of the wedge with a a visco-elastic layer, which is thin compared to the thickness of the plate (which here is allowed to be non-constant), i.e. $\delta/h(\xi) \ll h_1$, one can derive a formula for the wave numbers, where $\eta$ is simply replaced by 
\begin{align*}
\eta + \frac{3}{2}\frac{\delta}{h(\xi)h_1}\frac{E_1}{E_0}\tilde{\eta}.
\end{align*}
Following the same first order expansion as above, the imaginary part of the physical wave numbers are:
\begin{align}\label{eq:imwavenumberdampening}
\Im (k_\pm ) \approx\Big(\frac{\omega^2h_1^2}{8c_0^2 \kappa} \Big)^{1/2} \frac{\kappa^+ \pm \frac{2(\kappa^-)^2 + \tilde{h}_\xi}{2\sqrt{(\kappa^-)^2 + \tilde{h}_\xi}}}{\sqrt{\kappa^{+} \pm \sqrt{(\kappa^-)^2 + \tilde{h}_\xi}}} \Big[ \eta + \frac{3}{2}\frac{\delta}{h(\xi)h_1}\frac{E_1}{E_0}\tilde{\eta}\Big], \quad\kappa^+ \pm \sqrt{(\kappa^-)^2+\tilde{h}_\xi}>0.
\end{align}
An interesting open question here would be to consider how one could get a better performing acoustic black hole by including the thickness $\delta$ of the visco-elastic layer in the optimization problem by allowing it to vary as a function of $\xi$, but still under the condition that $\delta/h(\xi) \ll h_1$. We will not address this issue here. 
%{\color{red}
%and
%\begin{align*}
%\Im(k) = \Big(\frac{\omega^2h_1^2\rho}{2E_0 \kappa} \Big)^{1/2} \sqrt{\sqrt{(\kappa^-)^2+\tilde{h}_\xi}-\kappa^+}\Big[ \eta + \frac{3}{2}\frac{\delta}{h(\xi)}\frac{E_2}{E_1}\tilde{\eta}\Big], \quad \textup{ when } \kappa^+ - \sqrt{(\kappa^-)^2+\tilde{h}_\xi}<0.
%\end{align*}}

%
%
%
%
% APPROACH
%
%
%
%
\subsection{Lagrange optimization inspired method}\label{sec:lagrangemethod}
The problem of maximizing the integral \eqref{eq:integralrecflectioncoeff} under a pointwise constraint like in \eqref{eq:NWV} is quite complicated, for which no general recipe is available. Instead of a pointwise constraint, we will impose a weaker integral condition,  implemented by a penalty term, which guarantees at least that \eqref{eq:NWV} cannot be violated on a too large interval in $\xi$. Our method is related  to the Lagrange multiplier method from calculus of variation (see  \cite{GelfandFomin2012} and \cite{ShamesDym2003}).

More precisely, given some $n\geq 1$ and $\beta>0$, we will maximize the functional
\begin{align}\label{eq:integralfunctional}
I (h) = \int_0^t \Im(k(\xi))\d\xi -  \int_0^t\Big( \beta^{-1} \Big\vert \frac{1}{k^2}\frac{\d k}{\d \xi}\Big\vert \Big)^{2n} \d \xi,
\end{align}
over $C^1$ dimensionless, non-negative and non-decreasing height functions $h$ which obey $h(0)=h_\ell$ and $h(1)=1$. The maximizer of such a functional should obey 
\begin{align*}
 \Big\vert \frac{1}{k^2}\frac{\d k}{\d \xi} \Big\vert \leq \beta 
\end{align*}
on a large piece of the interval $[0,t]$, larger as $n$ grows, otherwise the penalty term would decrease the value of the functional.    This penalty is similar to the one introduced in topology optimization \cite{SigmundMaute}.

\subsection{Euler-Lagrange equation}\label{sec:eulerlagrange}
A great advantage coming from working with \eqref{eq:integralfunctional} is that we can rewrite the integrand in terms of $h(\xi)$ and $h'(\xi)$ and identify an Euler-Lagrange equation for $h$. The integrand does not explicitly depend on $\xi$, which makes that the "energy" is conserved and we can reduce the problem to  a first order equation. Using the expression from \eqref{hc3} of $k(\xi)$ where we neglect the imaginary part after differentiation, by elementary calculations it follows that 
\begin{align*}
\Bigg\vert\frac{1}{k^2}\frac{\mathrm{d}k}{\mathrm{d}\xi}\Bigg\vert^{2n} &\approx   h'(\xi)^{2n} \frac{1}{4^{2n}}\Big(\frac{2c_o^2\kappa}{\omega^2 h_1^2}\Big)^{n} \Bigg\vert\kappa^+ \pm \sqrt{(\kappa^-)^2+\tilde{h}_\xi}\Bigg\vert^{-3n}   \Big((\kappa^-)^2+\tilde{h}_\xi\Big)^{-n}  \Bigg(\frac{96\kappa^2c_0^2}{(h(\xi))^3\omega^2h_1^2}\Bigg)^{2n}.
\end{align*}
%\begin{align*}
%\frac{\d k}{\d \xi} &={\color{red}-}\Big(\frac{24h_1E_0^{1/2}\kappa^{3/2}}{\sqrt{2}\rho^{1/2}\omega}\Big) \frac{1}{\sqrt{\kappa^{+}(1+i\eta) \pm \sqrt{(\kappa^{-})^2 + \tilde{h}_\xi + i \eta \Big(2(\kappa^{-})^2 + \tilde{h}_\xi \Big)}}} \\
%&\quad \cdot \frac{\pm 1}{\sqrt{(\kappa^{-})^2+h_\xi+i\eta(2(\kappa^{-})^2+\tilde{h}_\xi)}}\frac{h'(\xi)(1+i \eta)}{(h(\xi))^3} 
%\end{align*}
%and
%\begin{align*}
%\frac{1}{k^2}=\Big(\frac{2E_0\kappa}{\omega^2h_1^2\rho}\Big) \frac{1}{\kappa^+(1+i\eta) \pm \sqrt{\kappa^-+\tilde{h}_\xi + i \eta E_0 \Big(2\kappa^- + \tilde{h}_\xi \Big)}}.
%\end{align*}
%Then, by using a first order Taylor approximation and the approximation that $\eta^2 \approx 0$, we get
%\begin{align*}
%\Big\vert \frac{1}{k^2}\frac{\mathrm{d}k}{\mathrm{d}\xi}\Big\vert &=\frac{12h_1^2\kappa}{E_0} \vert \kappa^+ \pm \sqrt{\kappa^- + \tilde{h}_\xi}\vert^{-3/2} \vert \kappa^- + \tilde{h}_\xi \vert^{-1/2} \frac{h'(\xi)}{(h(\xi))^3}
%\end{align*}

To ease notation in the following let $\tilde{h}(x)\coloneqq \frac{48\kappa^2c_0^2}{x^2\omega^2h_1^2}$. We define the function $\tilde{F}_\pm \colon (0,\infty) \times \mathbb{R} \to \mathbb{R}$ by
\begin{align*}
\tilde{F}_\pm(x,y) &\coloneqq \Big(\frac{\omega^2h_1^2}{8c_0^2 \kappa} \Big)^{1/2} \frac{\kappa^+ \pm \frac{2(\kappa^-)^2 + \tilde{h}(x)}{2\sqrt{(\kappa^-)^2 + \tilde{h}(x)}}}{\sqrt{\kappa^{+} \pm \sqrt{(\kappa^-)^2 + \tilde{h}(x)}}} \Big[ \eta + \frac{3}{2}\frac{\delta}{x h_1}\frac{E_1}{E_0}\tilde{\eta}\Big]\\
& - y^{2n} \beta^{-2n} \frac{1}{4^{2n}}\Big(\frac{2c_0^2\kappa}{\omega^2 h_1^2}\Big)^{n} \Bigg\vert\kappa^+ \pm \sqrt{(\kappa^-)^2+\tilde{h}(x)}\Bigg\vert^{-3n}   \Big((\kappa^-)^2+\tilde{h}(x)\Big)^{-n}  \Bigg(\frac{96\kappa^2c_0^2}{x^3\omega^2h_1^2}\Bigg)^{2n}.
\end{align*}
Our original functional can now be rewritten as 
$$\int_0^t \tilde{F}_\pm(h(\xi),h'(\xi))\d\xi.$$

Since the maximizer is invariant when the functional is  multiplied by a positive constant, 
we choose to divide our functional with all the constants on the second term of $\tilde{F}$  and consider the function $F_\pm \colon (0,\infty) \times \mathbb{R} \to \mathbb{R}$, given by 
\begin{align*}
F_\pm(x,y) &=b\frac{\kappa^+ \pm \frac{2(\kappa^-)^2 + \tilde{h}(x)}{2\sqrt{(\kappa^-)^2 + \tilde{h}(x)}}}{\sqrt{\kappa^{+} \pm \sqrt{(\kappa^-)^2 + \tilde{h}(x)}}} \Big[ \eta + \frac{3}{2}\frac{\delta}{xh_1}\frac{E_1}{E_0}\tilde{\eta}\Big]\\
&\qquad \qquad - y^{2n}\Bigg\vert\kappa^+ \pm \sqrt{(\kappa^-)^2+\tilde{h}(x)}\Bigg\vert^{-3n}   \Big((\kappa^-)^2+\tilde{h}(x)\Big)^{-n}  x^{-6n},
\end{align*}
where $b=2^{-(7n+3/2)}3^{-2n}\beta^{2n}\omega^{6n+1}h_1^{6n+1}(c_0^2)^{-(3n+1/2)}\kappa^{-(5n+1/2)}$.
From now on we will separately investigate the existence of a maximizer for the functionals
\begin{align*}
I_{\pm}(h)=\int_0^t F_\pm(h(\xi),h'(\xi)) d \xi. 
\end{align*}
Since $F_\pm$ do not depend explicitly on $\xi$, it follows by a standard result from Calculus of Variation (see eg. \cite{GelfandFomin2012}), that if a maximizer exists, then it must obey the first order initial value problem given by:
\begin{align*}
F_\pm(h(\xi),h'(\xi)) - h'(\xi) (\partial_y F_\pm)(h(\xi),h'(\xi)) = a,\quad h(t)=1,
\end{align*}
where $a$ is some constant which has to be chosen so that a $C^1$ non-decreasing solution exists on $(0,t)$ and which obeys $h(0)=h_\ell$.  We have:
\begin{align*}
\frac{\partial}{\partial y}F_\pm(x,y) = -2n y^{2n-1} \Bigg(\kappa^+ \pm \sqrt{(\kappa^-)^2+\tilde{h}(x)}\Bigg)^{-3n}   \Big((\kappa^-)^2+\tilde{h}(x)\Big)^{-n}  x^{-6n}.
\end{align*}
Isolating $h'(\xi)$ and asking that $h'\geq 0$, we obtain a  separable differential equation:
\begin{align}\label{hc4}
h'(\xi)&=\Bigg\{ \Bigg[a - b \frac{\kappa^+ \pm \frac{2(\kappa^-)^2 + \tilde{h}_\xi}{2\sqrt{(\kappa^-)^2 + \tilde{h}_\xi}}}{\sqrt{\kappa^{+} \pm \sqrt{(\kappa^-)^2 + \tilde{h}_\xi}}} \Big[ \eta + \frac{3}{2}\frac{\delta}{h(\xi)h_1}\frac{E_1}{E_0}\tilde{\eta}\Big]\Bigg]\nonumber \\
& \quad \times  \Bigg[(2n-1)  \Bigg\vert\kappa^+ \pm \sqrt{(\kappa^-)^2+\tilde{h}_\xi}\Bigg\vert^{-3n}   \Big((\kappa^-)^2+\tilde{h}_\xi\Big)^{-n}  (h(\xi))^{-6n}\Bigg]^{-1}\Bigg\}^{1/2n}.
\end{align}

We have not been able to explicitly solve this differential equation and we had to appeal to "semi-analytical" methods coupled with numerical tools, described in what follows. First, remember that $\tilde{h}_\xi=\frac{48\kappa^2c_0^2}{h(\xi)^2\omega^2h_1^2}$. Second, we notice that the right-hand side of \eqref{hc4} can be seen as a function $g_a:(0,\infty)\mapsto (0,\infty)$ of  $h(\xi)$, which also depends parametrically on $a$. Hence \eqref{hc4} can be rewritten as
$$h'(\xi)= g_a(h(\xi)),\quad h(t)=1.$$
Any local solution near $\xi=t$ must obey the functional equation
\begin{equation}\label{hc5}
 t-\xi=\int_{h(\xi)}^1 \frac{1}{g_a(y)}\d y.
 \end{equation}
If we start by choosing a very large $a$, then $1/g_a$ stays bounded on the whole interval $y\in [h_\ell,1]$ and we may compute the map 
$$a\mapsto \Phi(a):=\int_{h_\ell}^1 \frac{1}{g_a(y)}\d y$$
where the right-hand side goes to zero when $a\to\infty$. This means that if $a$ is large enough we have $t>\Phi(a)$, thus no such $a$ can insure the other boundary condition $h(0)=h_\ell$. 

Third, by reducing the value of $a$, $\Phi(a)$ increases continuously and there can be {\it at most} one value $a_*>0$ for which 
$\frac{1}{g_{a_*}(y)}$ is still integrable on $[h_\ell,1]$ and also $\Phi(a_*)=t$. With this value of $a=a_*$ we go back to \eqref{hc5} and find the unique value of $x\in [h_\ell,1]$ for which $\int_{x}^1 \frac{1}{g_{a_*}(y)}\d y$ equals $t-\xi$ for every $\xi\in [0,t]$.  

For some high frequency values, another phenomenon can appear. It might happen that by decreasing $a$ we arrive to a critical value $\tilde{a}_*$ such that $\frac{1}{g_{a}(y)}$ is no longer well-defined/integrable for $a<\tilde{a}_*$, while $\Phi(\tilde{a}_*)$ is still less than $t$. In this case, the optimal profile hits the value $h_\ell$ earlier than at $\xi=0$, namely at the point 
$t-\Phi(\tilde{a}_*)$, and then stays constant equal to $h_\ell$ between this point and zero.

\subsection{The second Timoshenko wave in the large frequency regime}\label{hc20}

If $|\Omega|$ is large enough, we saw that $k_+$ becomes independent of the profile, while $k_-$ is given by \eqref{hc7}.  Hence the relevant question here is to optimize the profile in order to minimize the reflection coefficient of the second wave. 

Denote by 
\begin{equation}\label{hc8}
    X(u):=\frac{8\kappa(1+\nu)\omega^2h_1^2\rho}{E_0}-\frac{48\kappa^2}{u^2},\quad Y(u):=\frac{8\kappa(1+\nu)\omega^2h_1^2\rho}{E_0}\Big (\eta + \frac{3}{2}\frac{\delta}{uh_1}\frac{E_1}{E_0}\tilde{\eta}\Big).
\end{equation}
Then using \eqref{hc7} we have 
\begin{equation}\label{hc9}
    {\rm Im}(k_-(\xi))\approx |\kappa-2(1+\nu)|^{-1/2}\frac{Y(h(\xi))}{2\sqrt{X(h(\xi))}}.
\end{equation}
Also, by neglecting the imaginary part we have:
\begin{equation}\label{hc10}
    |k_-^{-2}(\xi) k_-(\xi)'|^2\approx |\kappa-2(1+\nu)|\; \frac{48^2 \kappa^4 |h'(\xi)|^2}{X(h(\xi))^3 h(\xi)^6}
\end{equation}
By reasoning like in the previous case, and working with $n=1$ in the functional, there must exist a constant $a$ such that 
$$|\kappa-2(1+\nu)|^{-1/2}\frac{Y}{2\sqrt{X}}+\beta^{-2}|\kappa-2(1+\nu)|\; \frac{48^2 \kappa^4 |h'|^2}{X^3 h^6}=a,$$
or put differently, any $C^1$ increasing solution must obey
$$h'(\xi)=g_a(h(\xi)),\quad h(t)=1,$$
where 
$$g_a(u):=\Big (a-|\kappa-2(1+\nu)|^{-1/2}\frac{Y(u)}{2\sqrt{X(u)}}\Big )^{1/2}\frac{\beta X(u)^{3/2} u^3}{48 \kappa^2|\kappa-2(1+\nu)|^{1/2}},\quad h_\ell \leq u\leq 1.$$
When $n>1$, the formula for $g_a$ becomes:
$$g_a(u):=(2n-1)^{1/(2n)}\Big (a-|\kappa-2(1+\nu)|^{-1/2}\frac{Y(u)}{2\sqrt{X(u)}}\Big )^{1/(2n)}\frac{\beta X(u)^{3/2} u^3}{48 \kappa^2|\kappa-2(1+\nu)|^{1/2}}.$$

\section{Numerics}\label{sec:numerics}
 \subsection{Comparison between Euler-Bernoulli and the first Timoshenko wave} The dimensionless parameters of the wedge (see Fig.~\ref{fig:nondimtrunc}) are as follows
 \begin{align*}
     t=14.4, \quad h_\ell=0.0016, \quad n=1,
 \end{align*}
and the material parameters are those of steel. The loss factor of the wedge is taken as $\eta=0.001$. The loss factor of the visco-elastic layer $\tilde{\eta}=0.25$, while $\delta/h_1=0.0012$ and $\beta_2=E_1/E_0=0.3$. Lastly we take $\beta=\sqrt{2}/5$ as our penalty parameter.  

For clarity, we here present the results in terms of both dimensional and dimensionless frequencies. As we have already explained in the introduction, working with these parameters makes that the second Timoshenko wave (the one with $k_-$) only appears at very high ultrasound frequencies, when the first Timoshenko wave number (the one with $k_+$) becomes independent of the profile. That is why the predictions of the Euler-Bernoulli theory, valid at low frequencies,  can only be compared with those of the first Timoshenko wave. 
\begin{figure}[!h]
	\begin{minipage}{0.49\textwidth}
		\centering
    \includegraphics[scale=0.5]{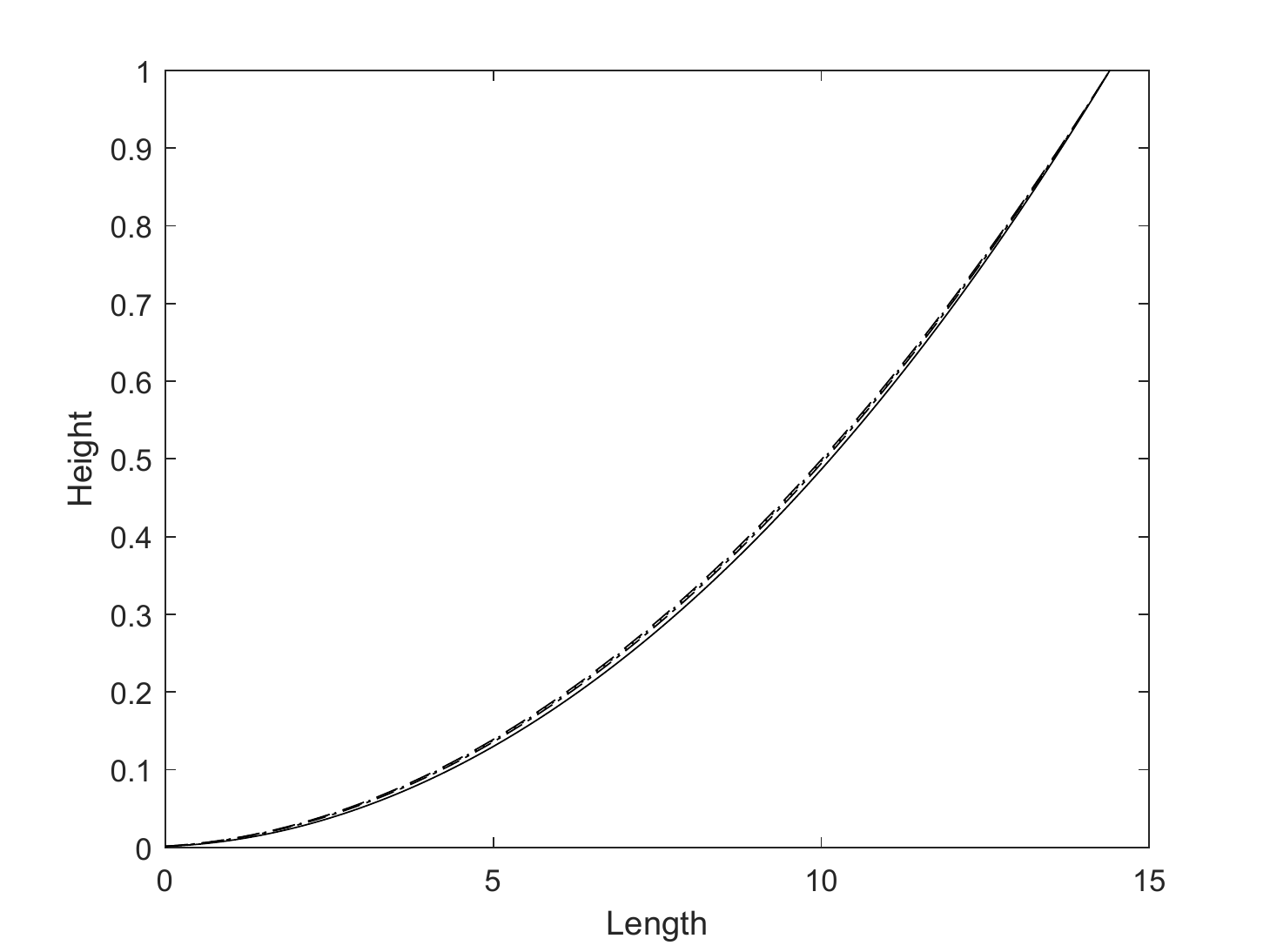}
    \caption{Profiles for $\Omega$ equal to $0.0025$ (dashed), $0.0124$ (dotted), $0.0248$ (dash-dotted), $0.0495$ (line), where both the length and height are dimensionless.}
    \label{fig:profile}
	\end{minipage}
	\begin{minipage}{0.49\textwidth}
		\centering
    \includegraphics[scale=0.5]{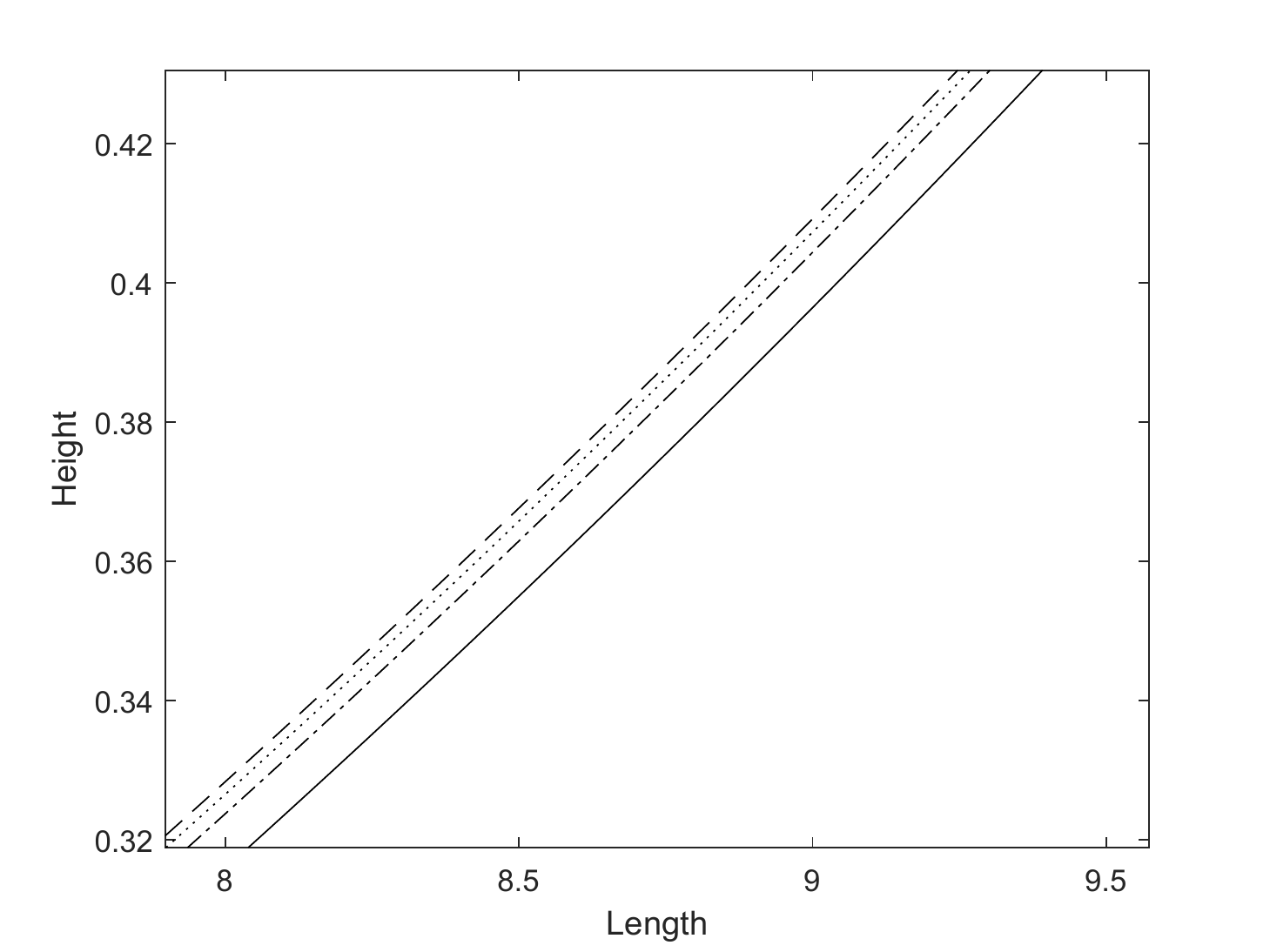}
    \caption{Zoomed in profiles for $\Omega$ equal to $0.0025$ (dashed), $0.0124$ (dotted), $0.0248$ (dash-dotted), $0.0495$ (line), where both the length and height are dimensionless.}
    \label{fig:profile_zoomedin}
	\end{minipage}
\end{figure}

\begin{figure}[!h]
	\begin{minipage}{0.49\textwidth}
		\centering
    \includegraphics[scale=0.16]{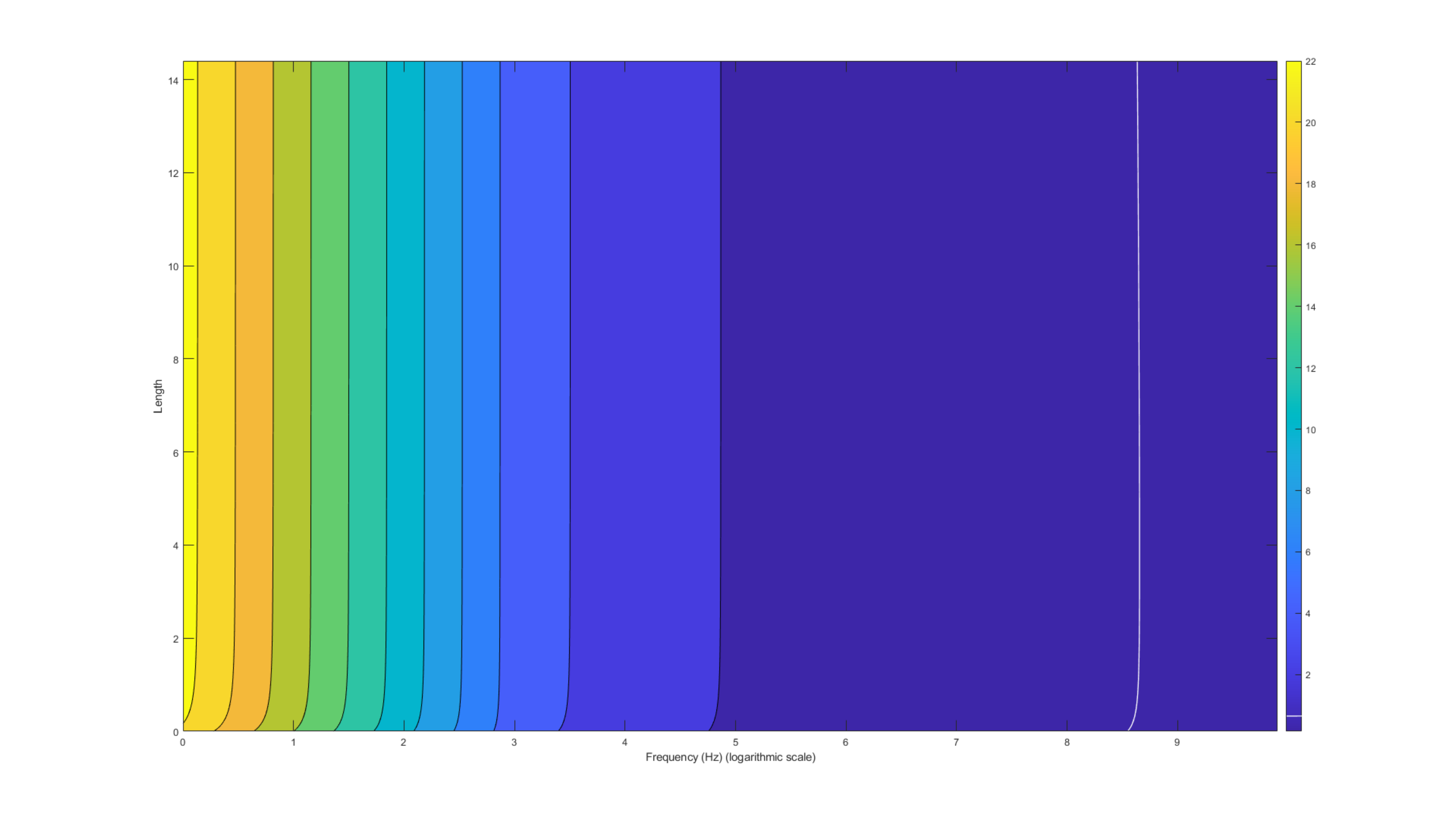}
    \caption{The NWV for optimal profile for $\Omega \approx 0.0025$.}
    \label{fig:NWVthousand}
	\end{minipage}
	\begin{minipage}{0.49\textwidth}
	    \centering
    \includegraphics[scale=0.16]{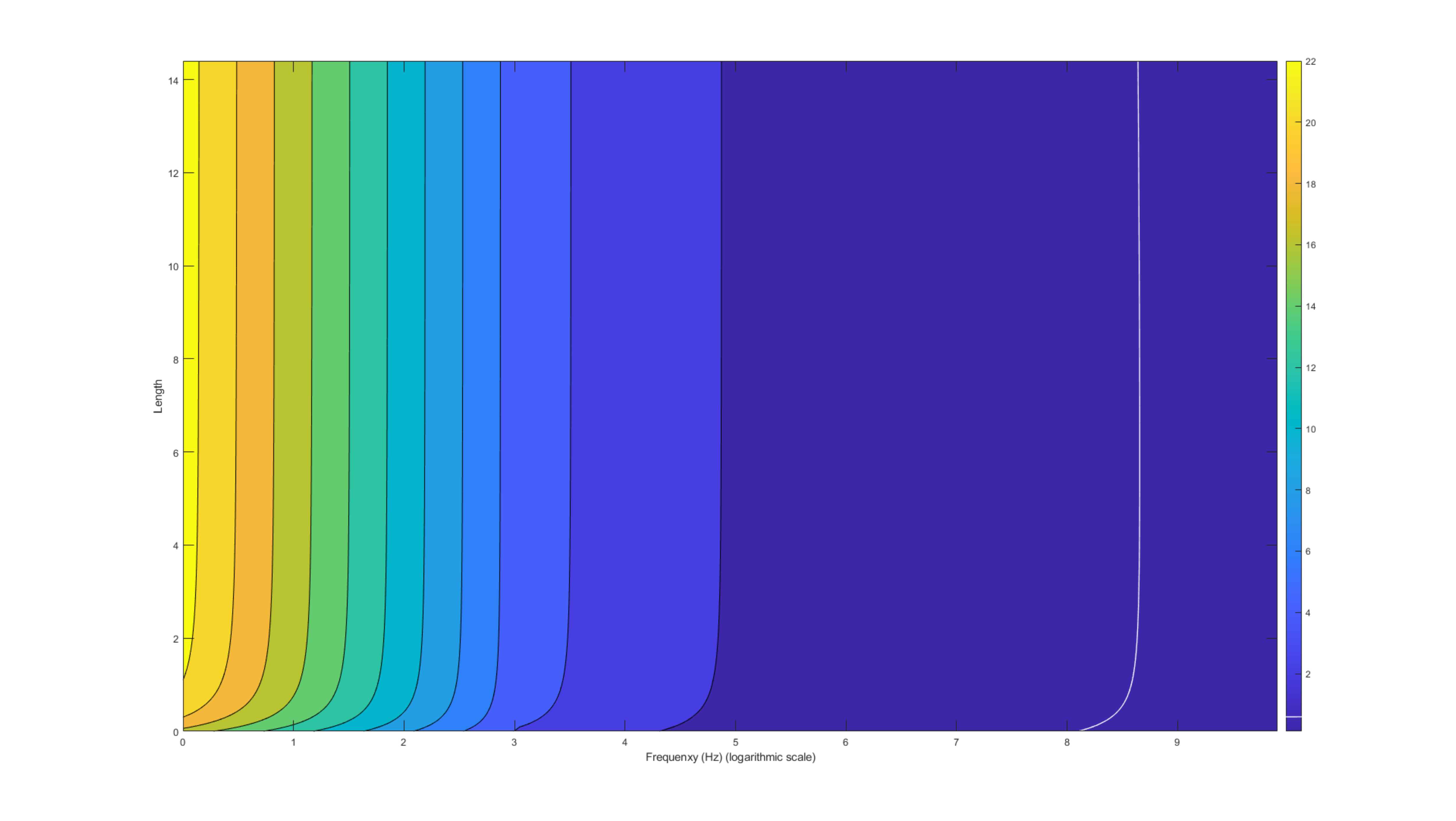}
    \caption{The NWV for optimal profile for $\Omega \approx 0.0124$.}
    \label{fig:NWVfivethousand}
	\end{minipage}
\end{figure}

\begin{figure}[!h]
\centering
    \includegraphics[scale=0.2]{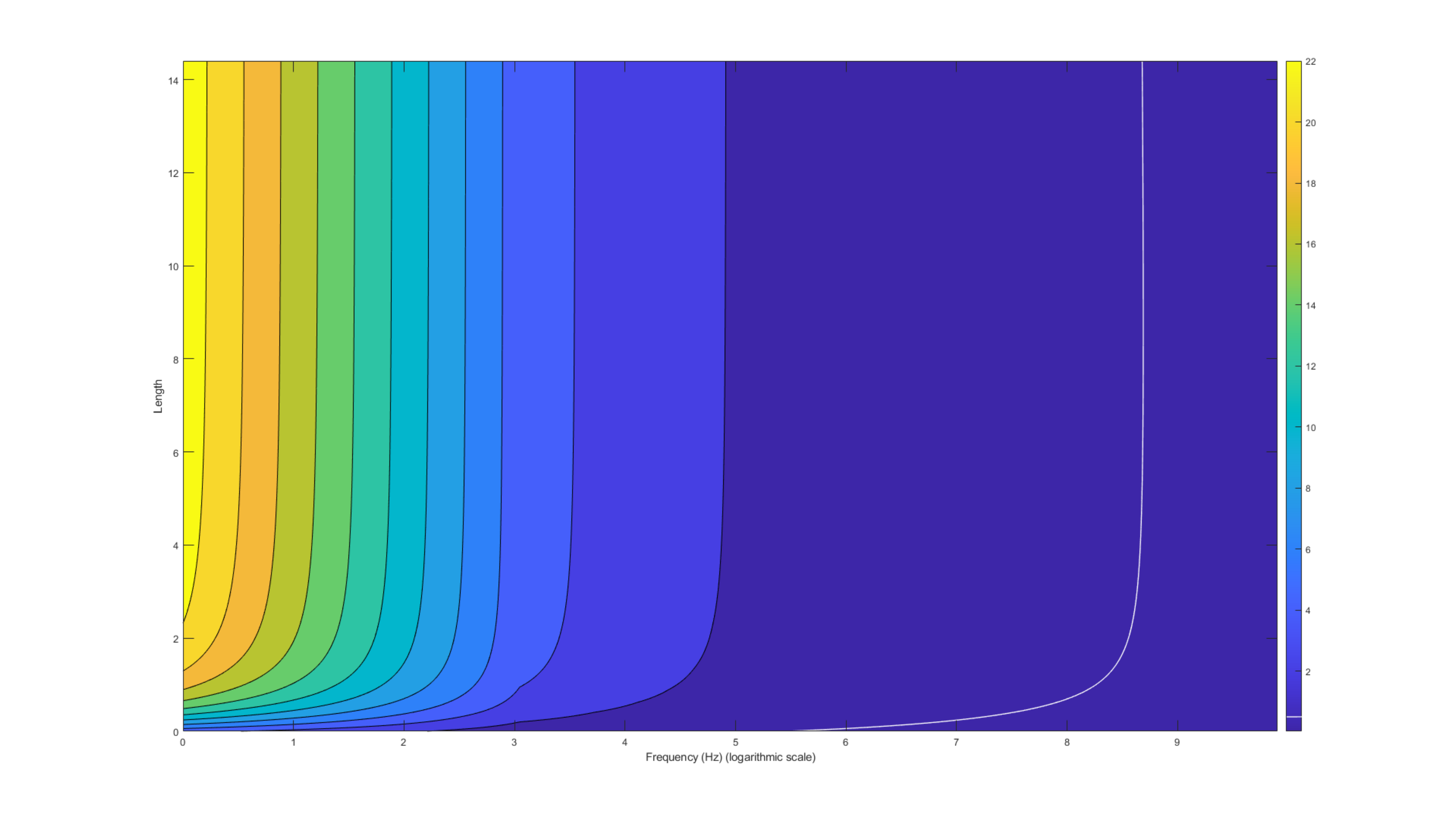}
    \caption{The NWV for optimal profile for $\Omega \approx 0.0495$}
    \label{fig:NWVtwentythousand}
\end{figure}

In Fig.~\ref{fig:profile} and Fig.~\ref{fig:profile_zoomedin}, we can see that the optimal profile  does not change much for $\Omega$ between $0.0025$ and $0.0495$ (1000Hz and 20000Hz) although the profile does get a little steeper as the frequency increases. The critical  behaviour seen in \cite{StottrupEtAl2021} for the Euler-Bernoulli beam theory, where the optimal profile has a constant part, also shows op in Timoshenko's framework, but for much larger frequencies. This is caused by the appearance of an $\tilde{a}_*$ as explained before. With the material parameters we have used here, one gets a constant part in the optimal profile for $\Omega \approx 0.06$ (24546Hz). For frequencies which lies in the range where the height profile found using Timoshenko's beam theory is smooth, the profile can be well approximated by a second order polynomial. More precisely,  the optimal profile for steel and $\Omega \approx 0.495$ (20000 HZ) is well approximated by the polynomial
\begin{align*}
    p(x)=0.004673x^2+0.001673x+0.003816.
\end{align*}
In the rest of this section we will refer many times to the {\it Normalized Wave number Variation} (NWV) given by the left-hand side of \eqref{eq:NWV}. In Fig.~\ref{fig:NWVthousand}-~\ref{fig:NWVtwentythousand} we have plotted the NWV for the optimal profile for the following values of $\Omega$, $0.0025$, $0.0124$ and $0.0495$, respectively (1000 Hz, 5000Hz and 20000Hz). The white graph is set at $NWV=0.3$, i.e. the upper bound for which the constraint for the NWV is satisfied. Similar for all three profiles is that the condition for the NWV is satisfied over the whole profile for frequencies above $5500$Hz, but one can also note that the constraint is satisfied for much lower frequencies in the low part of the profile, as seen i Fig.~\ref{fig:NWVtwentythousand}.

It is interesting to note, that the optimal profile we find for $\Omega$ below $0.0136$ (5500Hz) {\it does not} satisfy the pointwise NWV constraint all the way. This is due to the fact that we have only worked with $n=1$, which for low frequencies is not enough to transfer the integral penalty into a pointwise one. We cannot implement a high $n$ with our current numerical approach, since already at $n=5$, the constant $b=9.65\cdot 10^{-100}$, which numerically cancels out the imaginary part in our functional and creates serious instabilities. It would therefore be interesting to develop other numerical methods to find the optimal profile for lower frequencies and see how the optimal profile changes for these frequencies, if the penalty for violating he NWV constraint is harsher.

\begin{figure}[!h]
	\begin{minipage}{0.49\textwidth}
		\centering
    \includegraphics[scale=0.5]{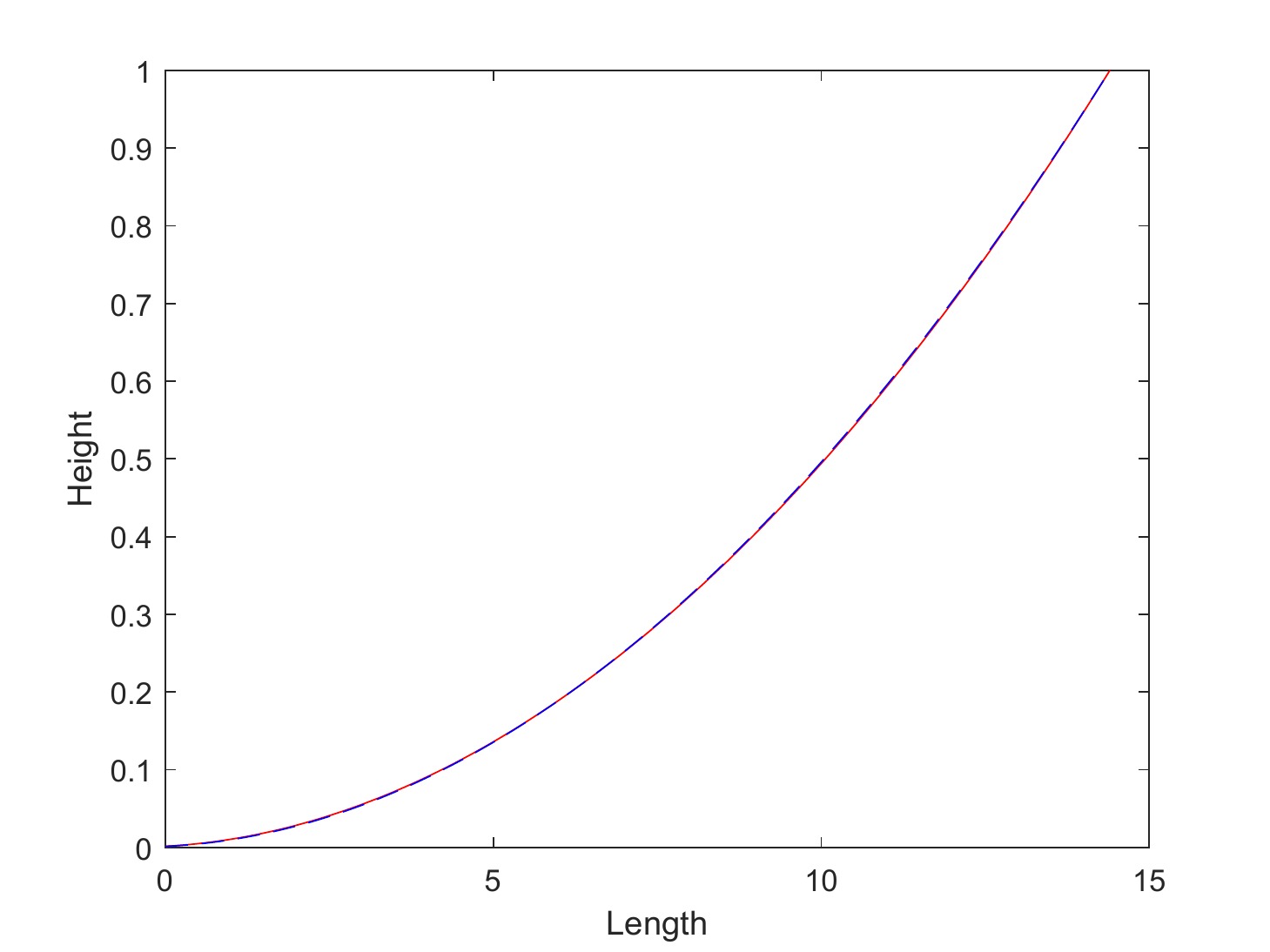}
     \caption{Wedgeprofile Euler-Bernoulli beam theory (blue) and Timoshenko beam theory (red) at $\Omega \approx 0.0248$, where both the length and height are dimensionless. }
    \label{fig:EBvsTimoprofileten}
	\end{minipage}
	\begin{minipage}{0.49\textwidth}
	    \centering
    \includegraphics[scale=0.5]{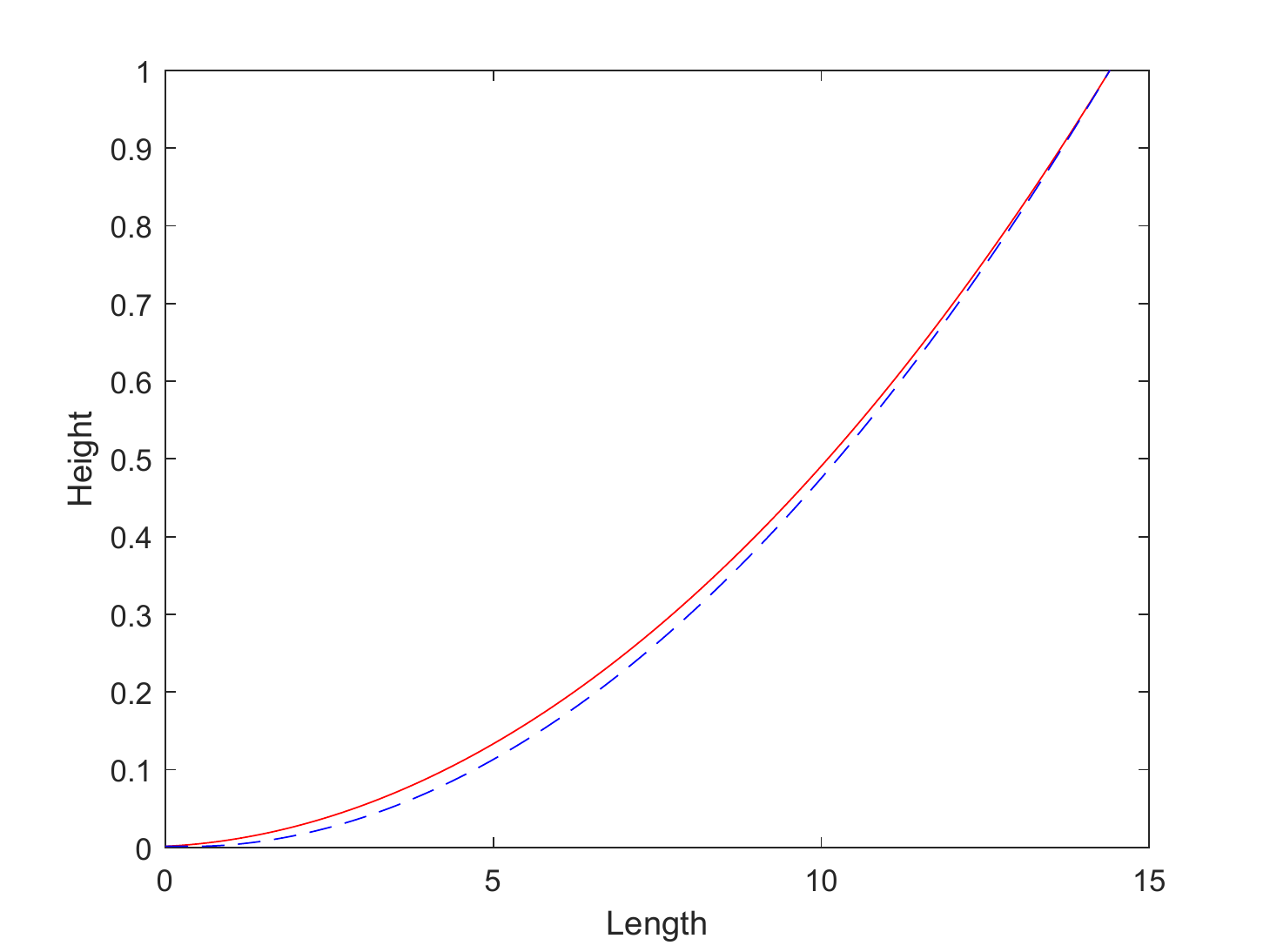}
    \caption{Wedge profile Euler-Bernoulli beam theory (blue) and Timoshenko beam theory (red) at $\Omega \approx 0.0371$, where both the length and height are dimensionless. }
    \label{fig:EBvsTimoprofilefifteen}
	\end{minipage}
\end{figure}

\begin{figure}[!h]
\centering
    \includegraphics[scale=0.5]{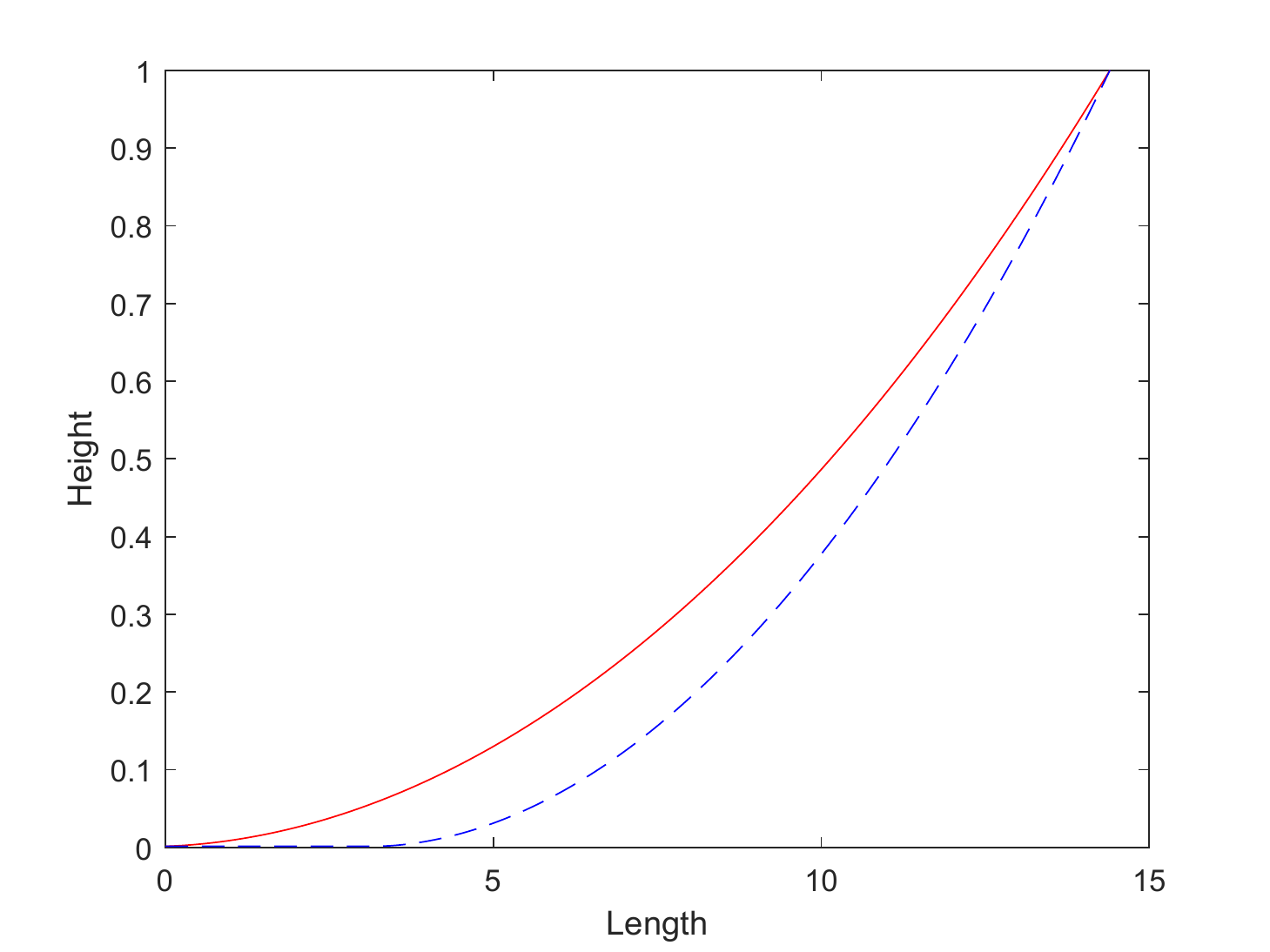}
    \caption{Wedge profile Euler-Bernoulli beam theory (blue) and Timoshenko beam theory (red) at $\Omega \approx 0.0495$, where both the length and height are dimensionless. }
    \label{fig:EBvsTimoprofiletwenty}
\end{figure}

In Fig.~\ref{fig:EBvsTimoprofileten}-~\ref{fig:EBvsTimoprofiletwenty} the optimal profile using both the Euler-Bernoulli and the Timoshenko beam theory is plotted for the following values of $\Omega$, $0.0248$, $0.0371$ and $0.0495$, respectively (10000Hz, 15000Hz and 20000Hz). Interestingly, the two profile coincides for frequencies up to $\Omega \approx 0.03$ (as seen eg. in Fig.~\ref{fig:EBvsTimoprofileten}) and then they start deviating from each other as seen in Fig.~\ref{fig:EBvsTimoprofilefifteen} and~\ref{fig:EBvsTimoprofiletwenty}, while the two theories agrees up to $\Omega \approx 0.3$

\subsection{Considerations on the second Timoshenko wave}

We have shown in Section \ref{hc20} that the second wave stops being evanescent and becomes relevant at frequencies for which the first wave becomes a Rayleigh wave. In other words, while the first wave can no longer be influenced by a clever choice of the profile, we can do something for the second wave. Mathematically, the optimal profiles for the second wave will have the same general features as for the first wave, which means that starting from some critically high frequency, the profiles will become piecewise constant near $\xi=0$, and also more and more steep. We decided not to add more plots because no qualitatively new phenomenon appears in this case.

\section{Conclusions and open questions}\label{sec:conclusion}
We have considered the problem of determining the optimal profile for a wedge in a beam  which minimizes the reflection coefficient of the first Timoshenko wave. This has been done within the mathematical framework of calculus of variations. We analytically derived an associated Euler-Lagrange equation which takes into account an integral penalty constraint on the normalized wave number variation (NWV), and then numerically solved this Euler-Lagrange equation in order to get the optimal profile. In the numerical investigation where we used the parameters of steel and a height ratio $h_\ell=0.0016$ we also show, that in the dimensionless frequency range $\Omega/h_\ell \ll 1$ where the Euler-Bernoulli and the Timoshenko beam theories certainly agree, the optimal profiles also agree. The two profiles start to differ at $\Omega > 0.0371$ (15000Hz). One interesting remark is that this change happens at $\Omega \approx 0.0371$ which is less than $0.3$, which is the threshold for which the two theories coincides for a plate with constant height. This shows that the transition between the two theories is more subtle in the variable height case.

Finally, we identified two open questions, which we believe it would be interesting to investigate in the near future:

\begin{itemize}
\item First, to try to find some material (or meta-material) where the second Timoshenko wave appears in the acoustic frequency range, and to numerically investigate whether the optimal profile for this wave compares to the one for the first wave.

\item Second, to find numerical methods for solving the Euler-Lagrange equation which better take into account the pointwise NWV constraint, like for example allowing a large $n$ in the integral penalty term. 

\end{itemize}

\bibliographystyle{plain}
\bibliography{bibliography}

\begin{thebibliography}{10}

\bibitem{Achenbach1973}
J.D. Achenbach.
\newblock {\em Wave Propagation in Elastic Solids}.
\newblock North-Holland, Amsterdam, 1973.

\bibitem{FeurtadoConlon}
P.~Feurtado and S.~Conlon.
\newblock Investigation of boundary-taper reflection for acoustic black hole
  design.
\newblock {\em J. Noise Cont. Eng.}, 63(5):460--466, 2015.

\bibitem{GelfandFomin2012}
I.M. Gelfand and S.V. Fomin.
\newblock {\em Calculus of Variations}.
\newblock Dover publications, 2000.

\bibitem{GuaschEtAl2017}
O.~Guasch, M.~Arnela, and P~Sanchez-Martin.
\newblock Transfer matrices to characterize linear and quadratic acoustic black
  holes in duct terminations.
\newblock {\em Journal of Sound and Vibration}, 395:65--79, 2017.

\bibitem{Krylov2004}
V.V Krylov.
\newblock New type of vibration dampers utilising the effect of acoustic 'black
  holes'.
\newblock {\em Acta Acust. United}, 90:830--837, 2004.

\bibitem{Krylov2007}
V.V. Krylov.
\newblock Propagation of plate bending waves in the vicinity of one- and
  two-dimensional acoustic black holes.
\newblock {\em ECCOMAS Thematic Conference on Computational Methods in
  Structural Dynamics and Earthquake}, pages 460--466, 2007.

\bibitem{KrylovTilman2004}
V.V. Krylov and F.J.B.S. Tilman.
\newblock Acoustic 'black holes' for flexural waves as effective vibration
  dampers.
\newblock {\em Journal of Sound and Vibration}, 274:605--619, 2004.

\bibitem{LeeJeon2017}
J.Y. Lee and W.~Jeon.
\newblock Vibration damping using a spiral acoustic black hole.
\newblock {\em The Journal of the Acoustical Society of America},
  141(3):1437--1445, 2017.

\bibitem{Miklovitz1980}
J.~Miklowitz.
\newblock {\em The Theory of Elastic Waves and Waveguides}.
\newblock North-Holland Publishing Company, 1980.

\bibitem{Mironov1988}
M.A. Mironov.
\newblock Propagation of a flexural wave in a plate whose thickness decreases
  smoothly to zero in a finite interval.
\newblock {\em Sov. Phys. Acoust.}, 34:318--319, 1988.

\bibitem{PelatEtAl2020}
A.~Pelat, F.~Gautier, S.C. Conlon, and F.~Semperlotti.
\newblock The acoustic black hole: A review of theory and applications.
\newblock {\em Journal of Sound and Vibration}, 476:1--24, 2020.

\bibitem{ShamesDym2003}
I.H. Shames and C.L. Dym.
\newblock {\em Energy and Finite Element Methods in Structural Mechanics}.
\newblock Taylor \& Francis, New York, 2003.

\bibitem{SigmundMaute}
O.~Sigmund and K.~Maute.
\newblock Topology optimization approaches.
\newblock {\em Struct. Multidic. Optim.}, 48:1031--1055, 2013.

\bibitem{SorokinChapman2011}
S.~Sorokin and C.J. Chapman.
\newblock A hierarchy of rational timoshenko dispersion relations.
\newblock {\em Journal of Sound and Vibration}, 330:5460--5473, 2011.

\bibitem{StephenPuchegger2006}
N.G. Stephen and S.~Puchegger.
\newblock On the valid frequency range of timoshenko beam theory.
\newblock {\em Journal of Sound and Vibration}, 297:1082--1087, 2006.

\bibitem{StottrupEtAl2021}
B.B. St{\o}ttrup, S.~Sorokin, and H.D. Cornean.
\newblock A rigorous approach to optimal profile design for acoustic black
  holes.
\newblock {\em The Journal of the Acoustical Society of America}, 149:447--456,
  2021.

\end{thebibliography}
\end{document}